\pdfoutput=1
\pdfoutput=1
\pdfoutput=1
\pdfoutput=1
\pdfoutput=1

\documentclass{article} % For LaTeX2e
\usepackage{iclr2017_conference,times}
\usepackage{url}

\usepackage{booktabs}       % professional-quality tables

\usepackage[pdftex]{graphicx}
\graphicspath{{./figure/}}
\usepackage{amsmath, amssymb}
\usepackage{cases}

\title{LSTM-Based System-Call Language Modeling
	\\ and Robust Ensemble Method for Designing\\
Host-Based Intrusion Detection Systems}

\author{Gyuwan Kim, Hayoon Yi, Jangho Lee, Yunheung Paek, Sungroh Yoon\thanks{To whom correspondence should be addressed.} \\
Seoul National University\\
\texttt{\{kgwmath,hyyi,ubuntu,ypaek,sryoon\}@snu.ac.kr} \\
}

\begin{document}

\maketitle

\begin{abstract}
  In computer security, designing a robust intrusion detection system is one of the most fundamental and important problems. In this paper, we propose a system-call language-modeling approach for designing anomaly-based host intrusion detection systems. To remedy the issue of high false-alarm rates commonly arising in conventional methods, we employ a novel ensemble method that blends multiple thresholding classifiers into a single one, making it possible to accumulate `highly normal' sequences. The proposed system-call language model has various advantages leveraged by the fact that it can learn the semantic meaning and interactions of each system call that existing methods cannot effectively consider. Through diverse experiments on public benchmark datasets, we demonstrate the validity and effectiveness of the proposed method. Moreover, we show that our model possesses high portability, which is one of the key aspects of realizing successful intrusion detection systems.
  %Besides, our model is computationally efficient in space and time.
\end{abstract}

\section{Introduction}
An intrusion detection system (IDS) refers to a hardware/software platform for monitoring network or system activities to detect malicious signs therefrom. Nowadays, practically all existing computer systems operate in a networked environment, which continuously makes them vulnerable to a variety of malicious activities. Over the years, the number of intrusion events is significantly increasing across the world, and intrusion detection systems have already become one of the most critical components in computer security. With the explosive growth of logging data, the role of machine learning in effective discrimination between malicious and benign system activities has never been more important.

A survey of existing IDS approaches needs a multidimensional consideration. Depending on the scope of intrusion monitoring, there exist two main types of intrusion detection systems: network-based (NIDS) and host-based (HIDS). The network-based intrusion detection systems monitor communications between hosts, while the host-based intrusion detection systems monitor the activity on a single system. From a methodological point of view, intrusion detection systems can also be classified into two classes~\citep{jyothsna2011}: signature-based and anomaly-based. The signature-based approaches match the observed behaviors against templates of known attack patterns, while the anomaly-based techniques compare the observed behaviors against an extensive baseline of normal behaviors constructed from prior knowledge, declaring each of anomalous activities to be an attack. The signature-based methods detect already known and learned attack patterns well but have an innate difficulty in detecting unfamiliar attack patterns. On the other hand, the anomaly-based methods can potentially detect previously unseen attacks but may suffer from making a robust baseline of normal behavior, often yielding high false alarm rates. The ability to detect a `zero-day' attack (i.e., vulnerability unknown to system developers) in a robust manner is becoming an important requirement of an anomaly-based approach. In terms of this two-dimensional taxonomy, we can classify our proposed method as an anomaly-based host intrusion detection system.

It was \citet{forrest1996} who first started to use system-call traces as the raw data for host-based anomaly intrusion detection systems, and system-call traces have been widely used for IDS research and development since their seminal work \citep{forrest2008}. System calls represent low-level interactions between programs and the kernel in the system, and many researchers consider system-call traces as the most accurate source useful for detecting intrusion in an anomaly-based HIDS. From a data acquisition point of view, system-call traces are easy to collect in a large quantity in real-time. Our approach described in this paper also utilizes system-call traces as input data.

For nearly two decades, various research has been conducted based on analyzing system-call traces. Most of the existing anomaly-based host intrusion detection methods typically aim to identify meaningful features using the frequency of individual calls and/or windowed patterns of calls from sequences of system calls. However, such methods have limited ability to capture call-level features and phrase-level features simultaneously. As will be detailed shortly, our approach tries to address this limitation by generating a language model of system calls that can jointly learn the semantics of individual system calls and their interactions (that can collectively represent a new meaning) appearing in call sequences.

In natural language processing (NLP), a language model represents a probability distribution over sequences of words, and language modeling has been a very important component of many NLP applications, including machine translation \citep{cho2014, bahdanau2014}, speech recognition \citep{graves2013}, question answering \citep{hermann2015}, and summarization \citep{rush2015}. Recently, deep recurrent neural network (RNN)-based language models are showing remarkable performance in various tasks~\citep{zaremba2014, jozefowicz2016}. It is expected that such neural language models will be applicable to not only NLP applications but also signal processing, bioinformatics, economic forecasting, and other tasks that require effective temporal modeling.

%There has been a long history of research to improve performance. Above all, recurrent neural network-based neural language models have shown remarkable performance \citep{zaremba2014, jozefowicz2016}. Moreover, this approach is applicable to any tasks involving sequential data, not only natural language processing but also signal processing, bioinformatics, economic forecasting, and other tasks.

Motivated by this performance advantage and versatility of deep RNN-based language modeling, we propose an application of neural language modeling to host-based introduction detection. We consider system-call sequences as a language used for communication between users (or programs) and the system. In this view, system calls and system-call sequences correspond to words and sentences in natural languages, respectively. Based on this system-call language model, we can perform various tasks that comprise our algorithm to detect anomalous system-call sequences: e.g., estimation of the relative likelihood of different words (i.e., system calls) and phrases (i.e., a window of system calls) in different contexts.

%We are motivated from this idea as well and extend it to the computer security domain. We can understand system call sequences as a language for communication between users (or programs) and the computer system. System calls and call sequences would correspond to characters and sentences in natural language, respectively. In that sense, we propose a system call language model, which is a probabilistic model that assigns a probability to a call sequence by predicting next calls in the sequence given a history of previous calls.

Our specific contributions can be summarized as follows: First, to model sequences of system calls, we propose a neural language modeling technique that utilizes long short-term memory (LSTM) \citep{hochreiter1997} units for enhanced long-range dependence learning. To the best of the authors' knowledge, the present work is the first end-to-end framework to model system-call sequences as a natural language for effectively detecting anomalous patterns therefrom.\footnote{In the literature, there exists only one related example of LSTM-based intrusion detection system~\citet{staudemeyer2013}, which, however, was in essence a feature-based supervised classifier (rather than an anomaly detector) requiring heavy annotation efforts to create labels. In addition, their approach was not an end-to-end framework and needed careful feature engineering to extract robust features for the classification.} Second, to reduce false-alarm rates of anomaly-based intrusion detection, we propose a  leaky rectified linear units (ReLU) \citep{maas2013rectifier} based ensemble method that constructs an integrative classifier using multiple (relatively weak) thresholding classifiers. Each of the component classifiers is trained to detect different types of `highly normal' sequences (i.e., system call sequences with very high probability of being normal), and our ensemble method blends them to produce a robust classifier that delivers significantly lower false-alarm rates than other commonly used ensemble methods.
As shown in Figure~\ref{fig:overview}, these two aspects of our contributions can seamlessly	be combined into a single framework. Note that the ensemble method we propose is not limited to our language-model based front-end but also applicable to other types of front-ends. %Also note that

In the rest of this paper, we will explain more details of our approach and then present our experimental results that demonstrate the effectiveness of our proposed method.

%The main point of this work is to introduce a neural language modeling framework to anomaly-based intrusion detection systems. We have selected long short-term memory (LSTM) \citep{hochreiter1997} as our basic unit, as it provides better learning capabilities for long-term sequence dependency. There has been previous work using LSTM in intrusion detection systems \citep{staudemeyer2013}, but it used hand-crafted features and carried out classification using labels rather than anomaly detection. Our method is an end-to-end approach with less prior knowledge. Our proposed model is extremely efficient and fast in both training and inference. Though it achieved quite excellent performance as it stands, we expect that future study will improve it even more.

%In addition, we propose a new ensemble method to build a single thresholding classifier by merging multiple thresholding classifiers. With the object of lowering false alarm rates, we adeptly gather `highly normal' sequences on the condition of not having many normal sequences and not having any abnormal sequences. The proposed ensemble method achieves significantly better performance compared to the results of other commonly known ensemble methods.

\section{Proposed Method}

\begin{figure}[t]
	\centering
	\includegraphics[width=\textwidth]{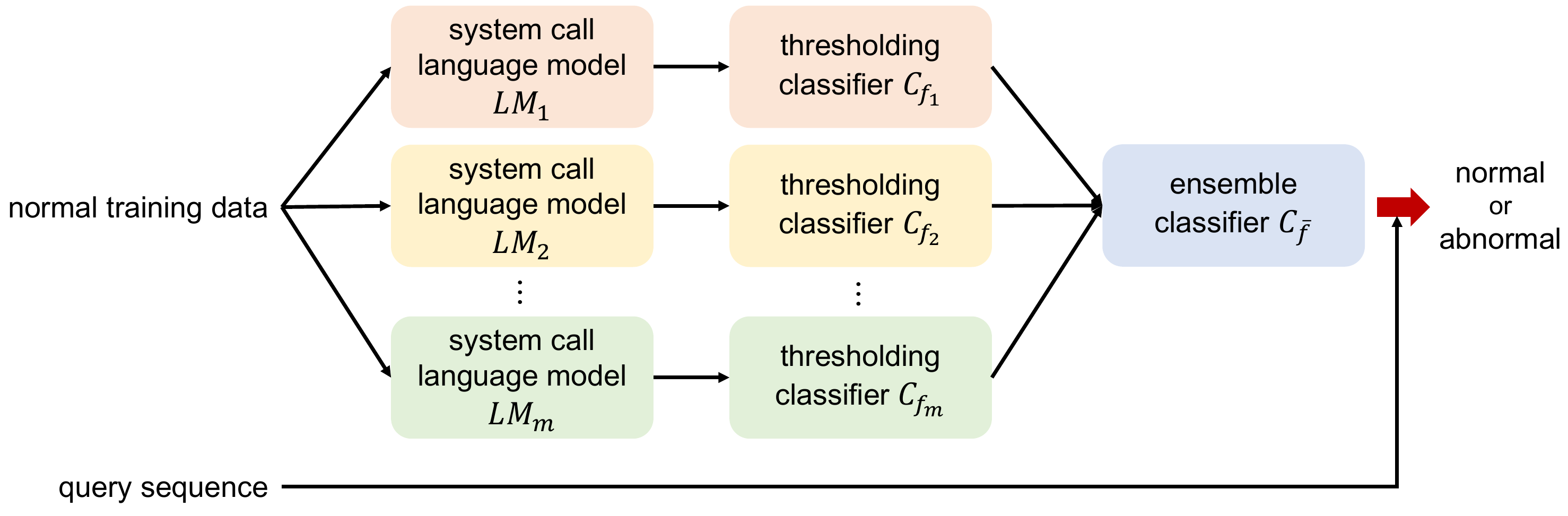}
	\caption{Overview of the proposed method.}
	\label{fig:overview}
\end{figure}

Figure~\ref{fig:overview} shows the overview of our proposed approach to designing an intrusion detection system. Our method consists of two parts: the front-end is for language modeling of system calls in various settings, and the back-end is for anomaly prediction based on an ensemble of thresholding classifiers derived from the front-end. In this section, we describe details of each component in our pipeline.

%Normal training data is used to train a system call language model. Multiple system call language models could be trained by changing hyper-parameters. For each system call language model, we can calculate a certain value corresponding to a sequence, for example, average negative log-likelihood and utilize this value for classification. By applying ensemble method, we could build a single classifier. This classifier will decide whether given query sequence is normal or abnormal. More details are explained in following subsections.

\subsection{Language Modeling of System Calls}
\begin{figure}[t]
  \centering
  \includegraphics[width=\textwidth]{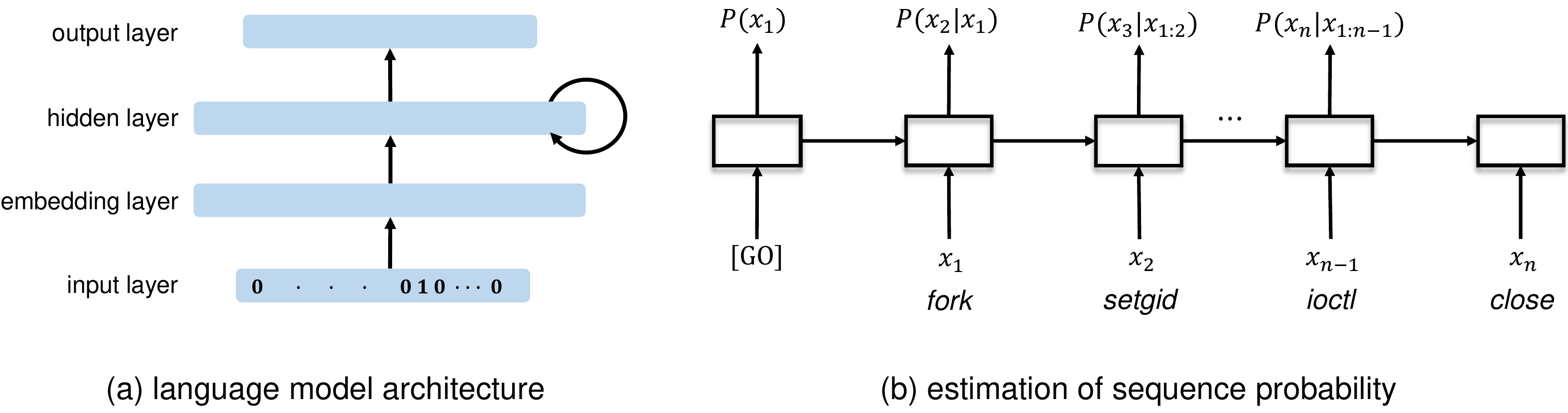}
  \caption{System-call language model.}
  \label{fig:architecture}
\end{figure}

Figure~\ref{fig:architecture} illustrates the architecture of our system-call language model. The system call language model estimates the probability distribution of the next call in a sequence given the sequence of previous calls. We assume that the host system generates a finite number of system calls. We index each system call by using an integer starting from $1$ and denote the fixed set of all possible system calls in the system as $S=\{1,\cdots,K\}$. Let $x=x_1x_2\cdots x_l (x_i\in S)$ denote a sequence of $l$ system calls.

At the input layer, the call at each time step $x_i$ is fed into the model in the form of one-hot encoding, in other words, a $K$ dimensional vector with all elements zero except position $x_i$. At the embedding layer, incoming calls are embedded to continuous space by multiplying embedding matrix $W$, which should be learned. At the hidden layer, the LSTM unit has an internal state, and this state is updated recurrently at each time step.
At the output layer, a softmax activation function is used to produce the estimation of normalized probability values of possible calls coming next in the sequence, $P(x_i|x_{1:i-1})$. According to the chain rule, we can estimate the sequence probability by the following formula:
\begin{align}
P(x)=\prod_{i=1}^{l}P(x_i|x_{1:i-1})
\end{align}

Given normal training system call sequence data, we can train this LSTM-based system call language model using the back-propagation through time (BPTT) algorithm. The training criterion minimizes the cross-entropy loss, which is equivalent to maximizing the likelihood of the system call sequence. A standard RNN often suffers from the vanishing/exploding gradient problem, and when training with BPTT, gradient values tend to blow up or vanish exponentially. This makes it difficult to learn long-term dependency in RNNs~\citep{bengio1994}. LSTM, a well-designed RNN architecture component, is equipped with an explicit memory cell and tends to be more effective to cope with this problem, resulting in numerous successes in recent RNN applications. %, as we can see in its name.

Because typical processes in the system execute a long chain of system calls, the number of system calls required to fully understand the meaning of a system-call sequence is quite large. In addition, the system calls comprising a process are intertwined with each other in a complicated way. The boundaries between system-call sequences are also vague. In this regard, learning long-term dependence is crucial for devising effective intrusion detection systems.

Markov chains and hidden Markov models are widely used probabilistic models that can estimate the probability of the next call given a sequence of previous calls. There has been previous work on using Markov models in intrusion detection systems \citep{hofmeyr1998, hoang2003, hu2009, yolacan2014}.
However, these methods have an inherent limitation in that the probability of the next call is decided by only a finite number of previous calls. Moreover, LSTM can model exponentially more complex functions than Markov models by using continuous space representations. This property alleviates the data sparsity issue that occurs when a large number of previous states are used in Markov models. In short, the advantages of LSTM models compared to Markov models are two folds: the ability to capture long-term dependency and enhanced expressive power.

Given a new query system-call sequence, on the assumption that abnormal call patterns deviate from learned normal patterns, yielding significantly lower probabilities than those of normal call patterns, a sequence with an average negative log-likelihood above a threshold is classified as abnormal, while a sequence with an average negative log-likelihood below the threshold is classified as normal. By changing the threshold value, we can draw a receiver operating characteristic (ROC) curve, which is the most widely used measure to evaluate intrusion detection systems.

Commonly, IDS is evaluated by the ROC curve rather than a single point corresponding to a specific threshold on the curve. Sensitivity to the threshold is shown on the curve. The x-axis of the curve represents false alarm rates, and the y-axis of the curve represents detection rates.\footnote{A false alarm rate is the ratio of validation normal data classified as abnormal. A detection rate is the ratio of detected attacks in the real attack data.} If the threshold is too low, the IDS is able to detect attacks well, but users would be annoyed due to false alarms. Conversely, if the threshold is too high, false alarm rates becomes lower, but it is easy for IDS to miss attacks. ROC curves closer to $(0,1)$ means a better classifier (i.e., a better intrusion detection system). The area under curve (AUC) summarizes the ROC curve into a single value in the range $[0,1]$ \citep{bradley1997}.

\subsection{Ensemble method to minimize false alarm rates}
\label{sec:ensemble}
Building a `strong normal' model (a model representing system-call sequences with high probabilities of being normal) is challenging because of over-fitting issues. In other words, a lower training loss does not necessarily imply better generalization performance. We can consider two reasons for encountering this issue.

First, it is possible that only normal data were used for training the IDS without any attack data. Learning discriminative features that can separate normal call sequences from abnormal sequences is thus hard without seeing any abnormal sequences beforehand. This is a common obstacle for almost every anomaly detection problem. In particular, malicious behaviors are frequently hidden and account for only a small part of all the system call sequences.

Second, in theory, we need a huge amount of data to cover all possible normal patterns to train the model satisfactorily. However, doing so is often impossible in a realistic situation because of the diverse and dynamic nature of system call patterns. Gathering live system-call data is harder than generating synthetic system-call data. The generation of normal training data in an off-line setting can create artifacts, because these data are made in fixed conditions for the sake of convenience in data generation. This setting may cause normal patterns to have some bias.

All these situations make it more difficult to choose a good set of hyper-parameters for LSTM architecture. To cope with this challenge, we propose a new ensemble method. Due to the lack of data, different models with different parameters capture slightly different normal patterns. If function $f\in S^{*}\mapsto \mathbb{R}$, which maps a system call sequence to a real value, is given, we can define a thresholding classifier as follows:
\begin{align}
C_{f}(x;\theta)=\begin{cases}
\textrm{normal} & \textrm{for} f(x)\leq\theta;\\
\textrm{abnormal} & \textrm{otherwise.} \end{cases}
\end{align}

Most of the intrusion detection algorithms, including our proposed method, employ a thresholding classifier. For the sake of explanation, we define a term  `highly normal' sequence for the classifier $C_f$ as a system call sequence having an extremely low $f$ value so it will be classified as normal even when the threshold $\theta$ is sufficiently low to discriminate true abnormals. Highly normal sequences are represented as a flat horizontal line near $(1,1)$ in the ROC curve. The more the classifier finds highly normal sequences, the longer this line is. Note that a highly normal sequence is closely related to the false alarm rate.

Our goal is to minimize the false alarm rate through the composition of multiple classifiers $C_{f_1}, C_{f_2}, \ldots, C_{f_m}$ into a single classifier $C_{\overline{f}}$, resulting in accumulated `highly normal' data (here $m$ is the number of classifiers used in the ensemble). This is due to the fact that a low false alarm rate is an important requisite in computer security, especially in intrusion detection systems. Our ensemble method can be represented by a simple formula:
\begin{align}
\overline{f}(x)=\sum_{i=1}^{m}w_i\sigma(f_i(x)-b_i).\label{eq3}
\end{align}

As activation function $\sigma$, we used a leaky ReLU function, namely $\sigma(x)=\max(x,0.001x)$. Intuitively, the activation function forces potential `highly normal' sequences having $f$ values lower than $b_i$ to keep their low $f$ values to the final $\overline{f}$ value. If we use the regular ReLU function instead, the degree of `highly normal' sequences could not be differentiated. We set the bias term $b_i$ to the median of $f$ values of the normal training data. In (\ref{eq3}), $w_i$ indicates the importance of each classifier $f_i$. Because we do not know the performance of each classifier before evaluation, we set $w_i$ to $1/m$. Mathematically, this appears to be a degenerated version of a one-layer neural network. The basic philosophy of the ensemble method is that when the classification results from various classifiers are slightly different, we can make a better decision by composing them well. Still, including bad classifiers could degrade the overall performance. By choosing classifiers carefully, we can achieve satisfactory results in practice, as will be shown in Section~\ref{sec:performance}.

\subsection{Baseline Classifiers}
Deep neural networks are an excellent representation learning method. We exploit the sequence representation learned from the final state vector of the LSTM layer after feeding all the sequences of calls. For comparison with our main classifier, we use two baseline classifiers that are commonly used for anomaly detection exploiting vectors corresponding to each sequence: $k$-nearest neighbor (kNN) and $k$-means clustering (kMC). Examples of previous work for mapping sequences into vectors of fixed-dimensional hand-crafted features include normalized frequency and tf-idf \citep{liao2002,xie2014}.

Let $T$ be a normal training set, and let $\mathrm{lstm}(x)$ denotes a learned representation of call sequence $x$ from the LSTM layer. kNN classifiers search for $k$ nearest neighbors in $T$ of query sequence $x$ on the embedded space and measure the minimum radius to cover them all. The minimum radius $g(x; k)$ is used to classify query sequence $x$. Alternatively, we can count the number of vectors within the fixed radius, $g(x; r)$. In this paper, we used the former. Because the computational cost of a kNN classifier is proportional to the size of $T$, using a kNN classifier would be intolerable when the normal training dataset becomes larger.
\begin{align}
%\begin{split}
%g(x; k) &= min\left\{r|\sum_{y\in T}\biggl[d(\mathrm{lstm}(x),\mathrm{lstm}(y))\leq r\biggr]\geq k\right\} \\
g(x; k) &= \min r \quad \mathrm{s.t.} \sum_{y\in T}\biggl[d(\mathrm{lstm}(x),\mathrm{lstm}(y))\leq r\biggr]\geq k \\
g(x; r) &= 1 - \frac{1}{|T|} \sum_{y\in T}\biggl[d(\mathrm{lstm}(x),\mathrm{lstm}(y))\leq r\biggr]
%\end{split}
\end{align}

The kMC algorithm partitions $T$ on the new vector space into $k$ clusters $G_1, G_2, \ldots, G_k$ in which each vector belongs to the cluster with the nearest mean so as to minimize the within-cluster sum of squares. They are computed by Lloyd's algorithm and converge quickly to a local optimum. The minimum distance from each center of clusters $\mu_i$, $h(x; k)$, is used to classify the new query sequence.
\begin{align}
h(x;k)=\min_{i=1,\cdots,k}d(\mathrm{lstm}(x),\mu_i)
\end{align}

The two classifiers $C_g$ and $C_h$ are closely related in that the kMC classifier is equivalent to the 1-nearest neighbor classifier on the set of centers. In both cases of kNN and kMC, we need to choose parameter $k$ empirically, depending on the distribution of vectors. In addition, we need to choose a distance metric on the embedding space; we used the Euclidean distance measure in our experiments.

\section{Experimental Results and Discussion}
\subsection{Datasets}
Though system call traces themselves might be easy to acquire, collecting or generating a sufficient amount of meaningful traces for the evaluation of intrusion detection systems is a nontrivial task. In order to aid researchers in this regard, the following datasets were made publicly available from prior work: ADFA-LD \citep{creech2013}, KDD98 \citep{lippmann2000} and UNM \citep{unm2004}. The KDD98 and UNM datasets were released in 1998 and 2004, respectively. Although these two received continued criticism about their applicability to modern systems \citep{brown2009, mchugh2000, tan2003}, we include them as the results would show how our model fares against early works in the field, which were mostly evaluated on these datasets. As the ADFA-LD dataset was generated around 2012 to reflect contemporary systems and attacks, we have done our evaluation mainly on this dataset.

The ADFA-LD dataset was captured on an x86 machine running Ubuntu 11.04 and consists of three groups: normal training traces, normal validation traces, and attack traces. The KDD98 dataset was audited on a Solaris 2.5.1 server. We processed the audit data into system call traces per session. Each session trace was marked as normal or attack depending on the information provided in the accompanied \texttt{bsm.list} file, which is available alongside the dataset. Among the UNM process set, we tested our model with \texttt{lpr} that was collected from SunOS 4.1.4 machines. We merged the live \texttt{lpr} set and the synthetic \texttt{lpr} set. This combined dataset is further categorized into two groups: normal traces and attack traces. To maintain consistency with ADFA-LD, we divided the normal data of KDD98 and UNM into training and validation data in a ratio of 1:5, which is the ratio of the ADFA-LD dataset. The numbers of system-call sequences in each dataset we used are summarized in Table~\ref{tab:dataset}.

%%%%% Add vocabulary size
\begin{table}[h]
  \caption{Summary of datasets used for experiments}
  \label{tab:dataset}
  \centering
  \begin{tabular}{lcrrcrr}
    \toprule
    %\hline
                  &\qquad\qquad & \multicolumn{2}{c}{\textbf{Normal}}  &\qquad\qquad& \multicolumn{2}{c}{\textbf{Attack}}  \\
                   %\cline{2-5}
     Benchmark             & & \# training & \# validation & & \# type    & \# attack \\
    \midrule
    ADFA-LD       & & 833         & 4372          & & 6          & 746            \\
    KDD98         & & 1364        & 5459          & & 10         & 41           \\
    UNM-lpr       & & 627         & 3136          & & 1          & 2002           \\
    %\hline
    \bottomrule
  \end{tabular}
\end{table}

\subsection{Performance Evaluation}
\label{sec:performance}
\begin{figure}[t]
  \centering
  \includegraphics[width=0.95\textwidth]{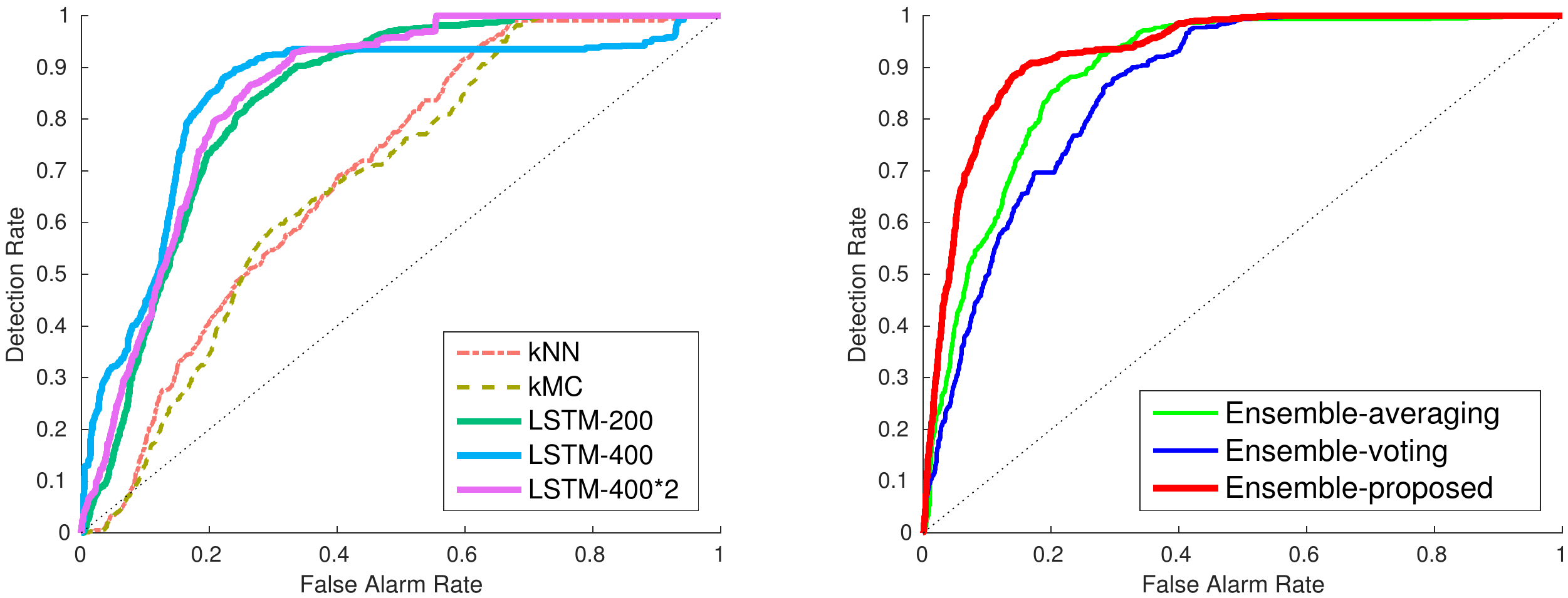}
  \caption{ROC curves from the ADFA-LD. Left shows the result from our three system-call language models with different parameters and two baseline classifiers. Right illustrates the results from different ensemble methods.}
  \label{fig:performance}
\end{figure}

We used ADFA-LD and built three independent system-call language models by changing the hyper-parameters of the LSTM layer: (1) one layer with $200$ cells, (2) one layer with $400$ cells, and (3) two layers with $400$ cells. We matched the number of cells and the dimension of the embedding vector. Our parameters were uniformly initialized in $[-0.1,0.1].$ For computational efficiency, we adjusted all system-call sequences in a mini-batch to be of similar lengths. We used the Adam optimizer \citep{kingma2014} for stochastic gradient descent with a learning rate of $0.0001$. The normalized gradient was rescaled whenever its norm exceeded 5 \citep{pascanu2013}, and we used dropout \citep{srivastava2014} with probability $0.5$. We show the ROC curves obtained from the experiment in Figure~\ref{fig:performance}.

For the two baseline classifiers, we used the Euclidean distance measure. Changing the distance measure to another metric did not perform well on average. In case of kNN, using $k=11$ achieved the best performance empirically. For kMC, using $k=1$ gave the best performance. Increasing the value of $k$ produced similar but poorer results. We speculate the reason why a single cluster suffices as follows: learned representation vectors of normal training sequence are symmetrically distributed. The kNN classifier $C_{g}$ and the kMC classifier $C_{h}$ achieved similar performance. Compared to \citet{liao2002,xie2014}, our baseline classifiers easily returned `highly normal' calls. This result was leveraged by the better representation obtained from the proposed system-call language modeling.

As shown in the left plot of Figure~\ref{fig:performance}, three LSTM classifiers performed better than $C_g$ and $C_h$. We assume that the three LSTM classifiers we trained are strong enough by themselves, and their classification results would be different from each other. By applying ensemble methods, we would expect to improve the performance. The first one was averaging, the second one was voting, and lastly we used our ensemble method as we explained in Section~\ref{sec:ensemble}. The proposed ensemble method gave a better AUC value ($0.928$) with a large margin than that of the averaging ensemble method ($0.890$) and the voting ensemble method ($0.859$). Moreover, the curve obtained from the proposed ensemble method was placed above individual single curves, while other ensemble methods did not show this property.

In the setting of anomaly detection where attack data are unavailable, learning ensemble parameters is infeasible. If we exploit partial attack data, the assumption breaks down and the zero-day attack issue remains. Our ensemble method is appealing in that it performs remarkably well without learning.

To be clear, we applied ensemble methods to three LSTM classifiers learned independently using different hyper-parameters, not with the baseline classifiers, $C_{g}$ or $C_{h}$. Applying ensemble methods to each type of baseline classifier gave unsatisfactory results since changing parameters or initialization did not result in complementary and reasonable classifiers that were essential for ensemble methods. Alternatively, we could do ensemble our LSTM classifiers and baseline classifiers together. However, this would also be a wrong idea because their $f$ values differ in scale. The value of $f$ in our LSTM classifier is an average negative log-likelihood, whereas $g$ and $h$ indicate distances in a continuous space.

According to \citet{creech2014}, the extreme learning machine (ELM) model, sequence time-delay embedding (STIDE), and the hidden Markov model (HMM) \citep{forrest1996, warrender1999} achieved about $13\%, 23\%, and 42\%$ false alarm rates (FAR) for $90\%$ detection rate (DR), respectively. We achieved $16\%$ FAR for $90\%$ DR, which is comparable result with the result of ELM and outperforms STIDE and HMM. The ROC curves for ELM, HMM, and STIDE can be found,  but we could not draw those curves on the same plot with ours because the authors provided no specific data on their results. \citet{creech2014} classified ELM as a semantic approach and other two as syntactic approaches which treat each call as a basic unit. To be fair, our proposed method should be compared with those approaches that use system calls only as a basic unit in that we watch the sequence call-by-call. Furthermore, our method is end-to-end while ELM relies on hand-crafted features.

\subsection{Portability Evaluation}
\begin{figure}[t]
  \centering
  \includegraphics[width=0.95\textwidth]{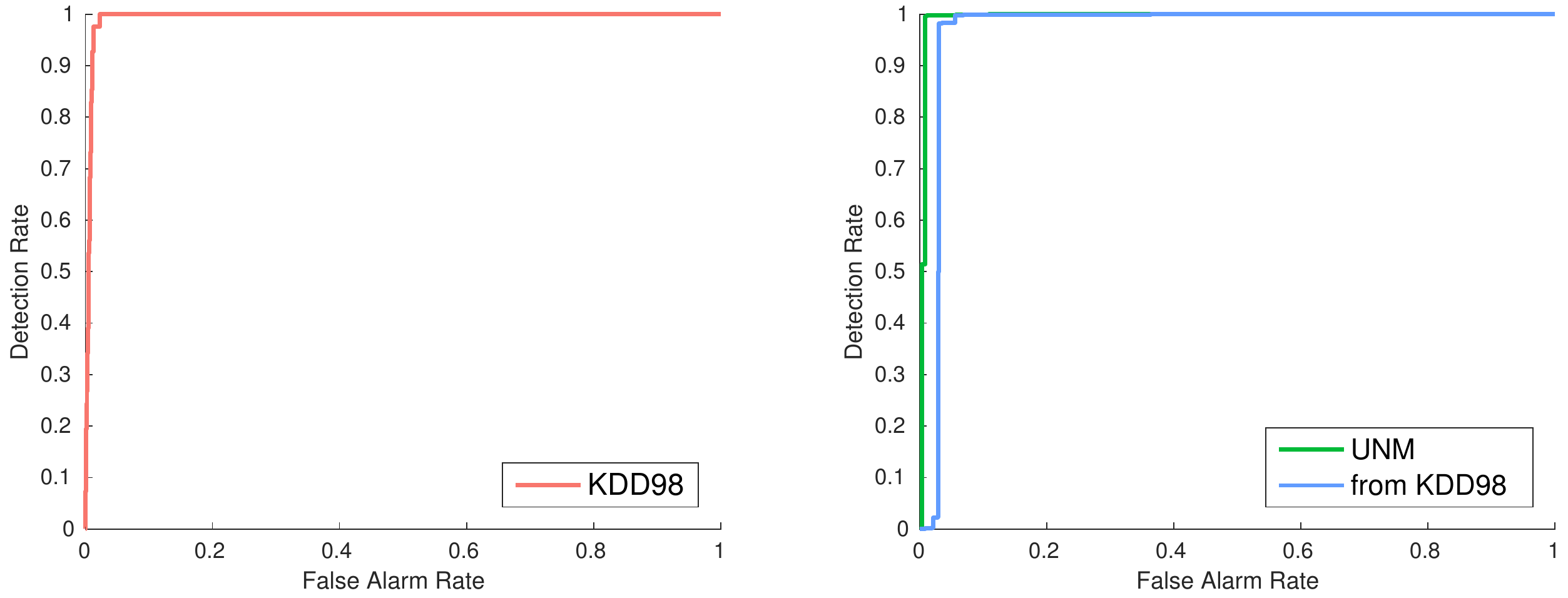}
  \caption{ROC curves from the KDD dataset and UNM dataset. Left is the evaluation about the KDD dataset. Right is the evaluation about UNM dataset using the model trained with the KDD98 dataset and the UNM dataset.}
  \label{fig:portability}
\end{figure}

We carried out experiments similar to those presented in Section~\ref{sec:performance} using the KDD98 dataset and the UNM dataset. First, we trained our system-call language model with LSTM having one layer of 200 cells and built our classifier using the normal training traces of the KDD98 dataset. The same model was used to evaluate the UNM dataset to examine the portability of the LSTM models trained with data from a different but similar system. The results of our experiments are represented in Figure~\ref{fig:portability}. For comparison, we display the ROC curve of the UNM dataset by using the model from training the normal traces therein. To examine portability, the system calls in test datasets need to be included or matched to those of training datasets. UNM was generated using an earlier version of OS than that of KDD98, but ADFA-LD was audited on a fairly different OS. This made our experiments with other combinations difficult.
%False alarm rate and detection rate were calculated by using the normal traces and attack traces respectively.

Through a quantitative analysis, for the KDD98 dataset, we earned an almost perfect ROC curve with an AUC value of $0.994$ and achieved $2.3\%$ FAR for $100\%$ DR. With the same model, we tested the UNM datset and obtained a ROC curve with an AUC value of $0.969$ and $5.5\%$ FAR for $99.8\%$ DR. This result was close to the result earned by using the model trained on normal training traces of the UNM dataset itself, as shown in the right plot of Figure~\ref{fig:portability}.

This result is intriguing because it indicates that system-call language models have a strong portability. In other words, after training one robust and extensive model, the model can then be deployed to other similar host systems. By doing so, we can mitigate the burden of training cost. This paradigm is closely related to the concept of transfer learning, or zero-shot learning. It is well known that neural networks can learn abstract features and that they can be used successfully for unseen data.
%universality

\subsection{Visualization of learned representations}
\begin{figure}[t]
  \centering
  \includegraphics[width=\textwidth]{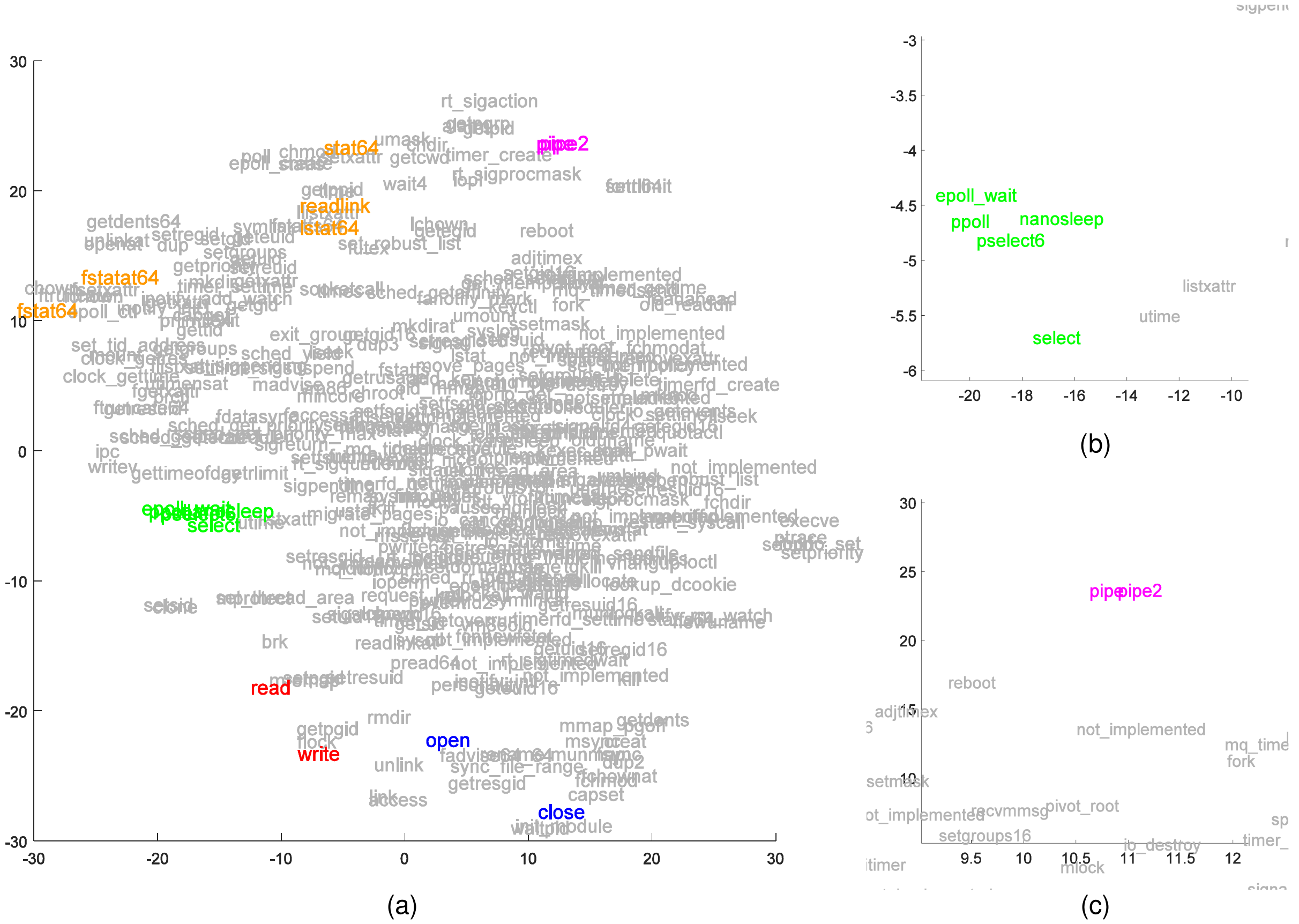}
  \caption{2-D embedding of learned call representations. (a) shows the full representation space of system calls that appeared in training data. (b) and (c) show the zoomed-in view of specific regions.}
  \label{fig:tsne_call}
\end{figure}

It is well-known that neural network based-language models can learn semantically meaningful embeddings to continuous space \citep{bengio2003, mikolov2013, cho2014}. We expected to see a similar characteristic with the proposed system-call language model. The 2–D projection of the calls using the embedding matrix $W$ learned from the system-call language model was done by t-SNE \citep{van2008} and shown in Figure ~\ref{fig:tsne_call}. Just as the natural language model, we can expect that calls having similar co-occurrence patterns are positioned in similar locations in the embedded space after training the system call language model. We can clearly see that calls having alike functionality are clustered with each other.

The first obvious cluster would be the \textit{read}-\textit{write} call pair and the \textit{open}-\textit{close} pair. The calls of each pair were located in close proximity in the space, meaning that our model learned to associate them together. At the same time, the difference between the calls of each pair appears to be almost the same in the space, which in turn would mean our model learned that the relationship of each pair somewhat resembles.

Another notable cluster would be the group of \textit{select}, \textit{pselect6}, \textit{ppoll}, \textit{epoll\_wait} and \textit{nanosleep}. The calls \textit{select}, \textit{pselect6} and \textit{ppoll} all have nearly identical functions in that they wait for some file descriptors to become ready for some class of I/O operation or for signals. The other two calls also have similar characteristics in that they wait for a certain event or signal as well. This could be interpreted as our model learning that these `waiting' calls share similar characteristics.

Other interesting groups would be: \textit{readlink} and \textit{lstat64} which are calls related to symbolic links; \textit{fstatat64} and \textit{fstat64} which are calls related to stat calls using file descriptors; \textit{pipe} and \textit{pipe2} which are nearly identical and appear almost as one on the embedding layer. These cases show that our model is capable of learning similar characteristics among the great many system calls.

Similarly to the call representations, we expected that attack sequences with the same type would cluster to each other, and we tried to visualize them. However, for various reasons including the lack of data, we were not able to observe this phenomenon. Taking the fact that detecting abnormal patterns from normal patterns well would be sufficiently hard into consideration, learning representation to separate different abnormal patterns with only seen normal patterns would also be an extremely difficult task.

\section{Conclusion}
Our main contributions for designing intrusion detection systems as described in this paper have two parts: the introduction of a system-call language modeling approach and a new ensemble method. To the best of the authors' knowledge, our method is the first to introduce the concept of a language model, especially using LSTM, to anomaly-based IDS. The system-call language model can capture the semantic meaning of each call and its relation to other system calls. Moreover, we proposed an innovative and simple ensemble method that can better fit to IDS design by focusing on lowering false alarm rates. We showed its outstanding performance by comparing it with existing state-of-the-art methods and demonstrated its robustness and generality by experiments on diverse benchmarks.

As discussed earlier, the proposed method also has excellent portability. In contrast to alternative methods, our proposed method incurs significant smaller training overhead because it does not need to build databases or dictionaries to keep a potentially exponential amount of patterns. Our method is compact and light in that the size of the space required to save parameters is small. The overall training and inference processes are also efficient and fast, as our methods can be implemented using efficient sequential matrix multiplications.

%We hope that this work will open a new approach in intrusion detection systems.

As part of our future work, we are planning to tackle the task of detecting elaborate contemporary attacks including mimicry attacks by more advanced methods. In addition, we are considering designing a new framework to build a robust model in on-line settings by collecting large-scale data generated from distributed environments. For optimization of the present work, we would be able to alter the structure of RNNs used in our system-call language model and ensemble algorithm. Finally, we anticipate that a hybrid method that combines signature-based approaches and feature engineering will allow us to create more accurate intrusion detection systems.

\subsubsection*{Acknowledgments}
%The authors would like to thank (P) for their constructive comments.
This work was supported by BK21 Plus Project in 2016 (Electrical and Computer Engineering, Seoul National University).

%\newpage
\bibliography{iclr2017_conference}
\bibliographystyle{iclr2017_conference}

\end{document}